\documentclass[sigconf, nonacm]{acmart}
\usepackage{caption}
\usepackage{subcaption}
\newcommand\vldbdoi{XX.XX/XXX.XX}

\newcommand\vldbavailabilityurl{URL_TO_YOUR_ARTIFACTS}

\newcommand{\fullindex}{Adaptive Hotspot-Aware Tree}
\newcommand{\abbrindex}{AHA-tree}
\newcommand{\bplustree}{B$^{+}$-tree}
\newcommand{\rtl}{\texttt{rootLSMT}}
\newcommand{\ndl}{\texttt{nodeLSMT}}

\begin{document}
\title{The AHA-Tree: An Adaptive Index for HTAP Workloads}

\author{Lu Xing}
\affiliation{
  \institution{Purdue University, West Lafayette, IN}
  \postcode{47907-2107}
}
\email{xingl@purdue.edu}

\author{Walid G. Aref}
\affiliation{%
  \institution{Purdue University, West Lafayette, IN}
  \postcode{47907-2107}
}
\email{aref@purdue.edu}

\begin{abstract}
In this demo, we realize data indexes that can morph from being write-optimized at times to being read-optimized at other times nonstop with zero-down time during the workload transitioning. These data indexes are useful for HTAP systems (Hybrid Transactional and Analytical Processing Systems), where transactional workloads are write-heavy while analytical workloads are read-heavy. Traditional indexes, e.g., \bplustree{} and LSM-Tree, although optimized for one kind of workload, cannot perform equally well under all workloads. To migrate from the write-optimized LSM-Tree to a read-optimized \bplustree{} is costly and mandates some system down time to reorganize data. We design adaptive indexes that can dynamically morph from a pure LSM-tree to a pure buffered B-tree back and forth, and has interesting states in-between. There are two challenges: allowing concurrent operations and avoiding system down time. This demo benchmarks the proposed AHA-Tree index under dynamic workloads and shows how the index evolves from one state to another without blocking.
\end{abstract}

\maketitle



\section{Introduction}\label{section:introduction}

\begin{figure}[h]
    \begin{subfigure}{\linewidth}
    \centering
        \includegraphics[width=\linewidth]{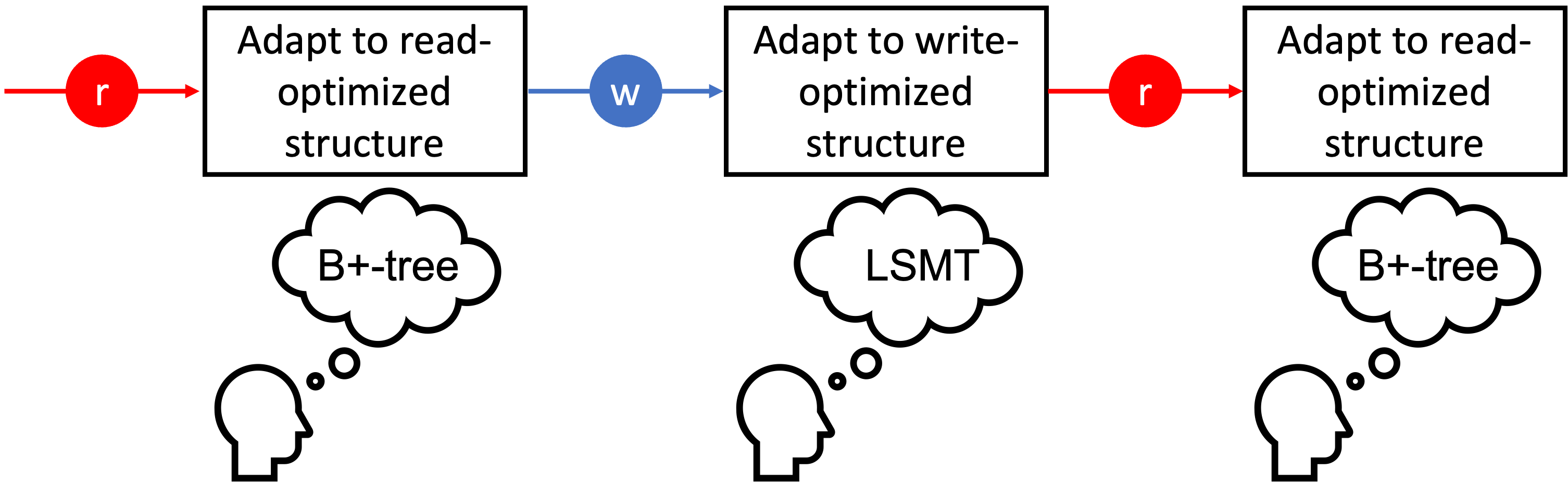}
        \caption{A diagram of an ideal index evolution under the oscillating read-heavy and write-heavy workloads}
        \label{fig:flow}
    \end{subfigure}
    \vfill
    \begin{subfigure}{\linewidth}
    \centering
        \includegraphics[width=\linewidth]{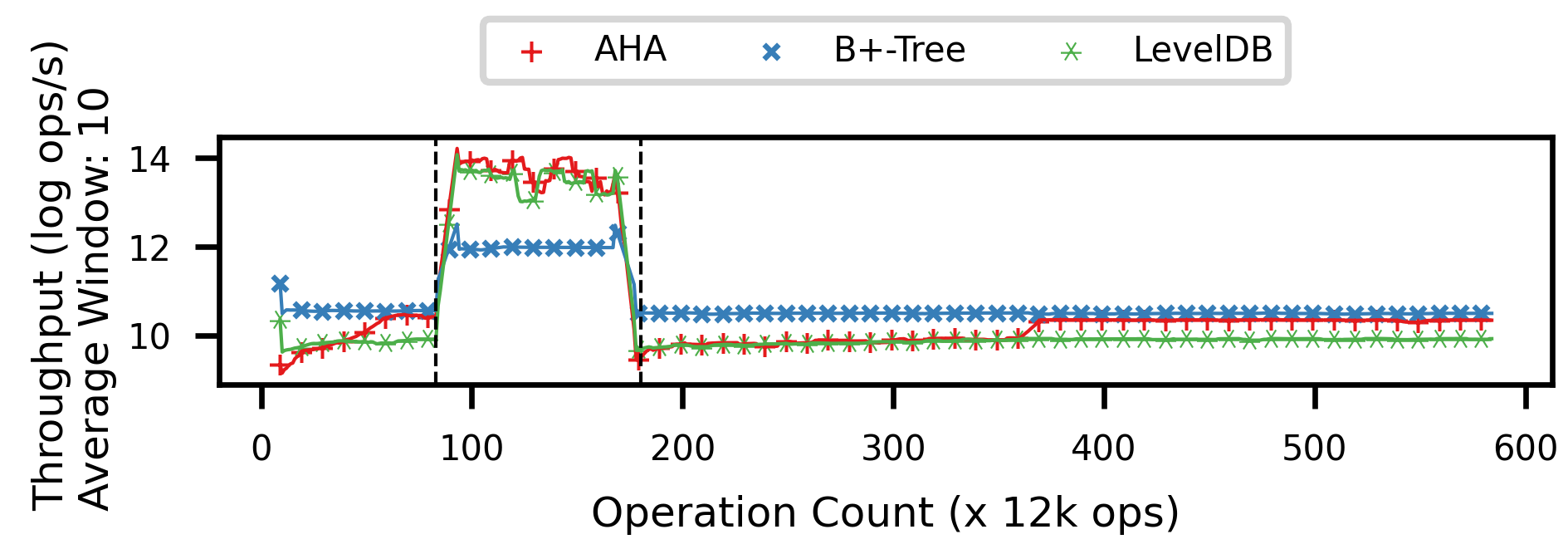}
        \caption{Throughput under a changing workload from read-heavy (0 to 1 million operation counts); write-heavy (1 to 2 million operation counts); read-heavy (remaining)}
        \label{fig:demo}
    \end{subfigure}
    \caption{The ideal and real performance of adaptive index and static index under a changing workload.}\label{fig:fig1}
\end{figure}

\begin{figure*}[h]
    \begin{subfigure}{0.49\linewidth}
    \centering
        \includegraphics[width=\linewidth]{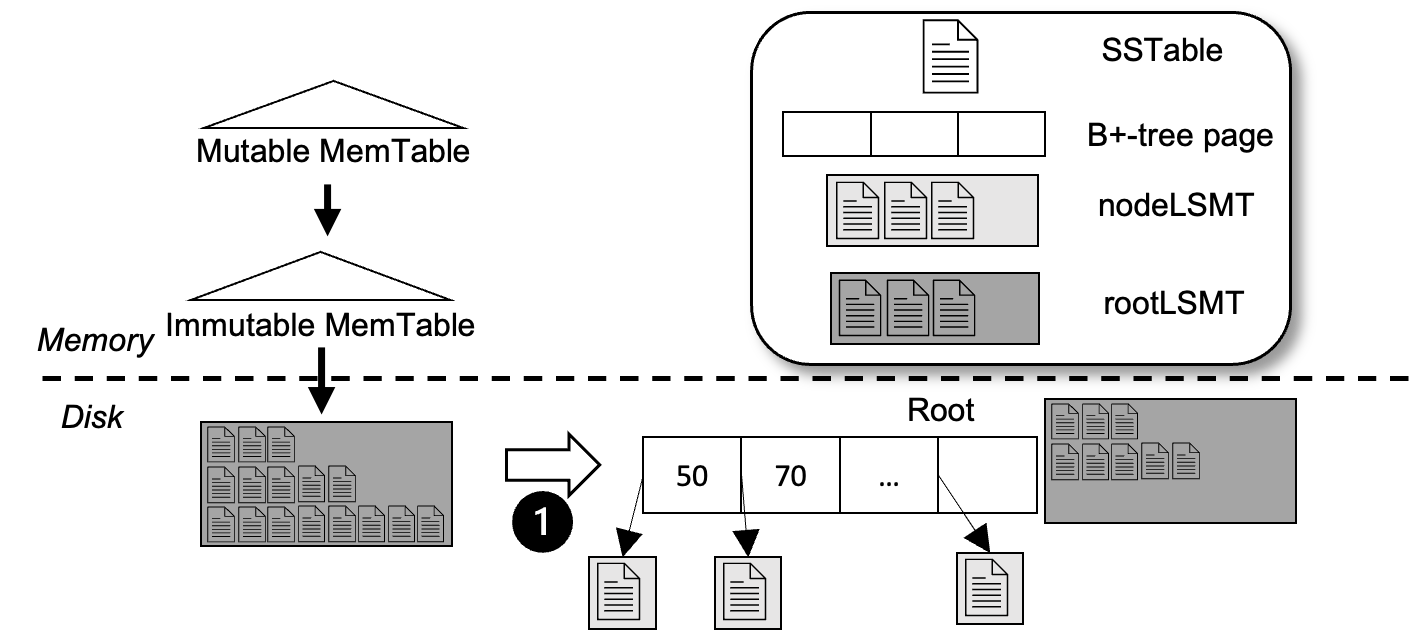}
        \caption{The insertion process in \abbrindex{}.}
        \label{fig:insert}
    \end{subfigure}
    \hfill
    \begin{subfigure}{0.49\linewidth}
    \centering
        \includegraphics[width=\linewidth]{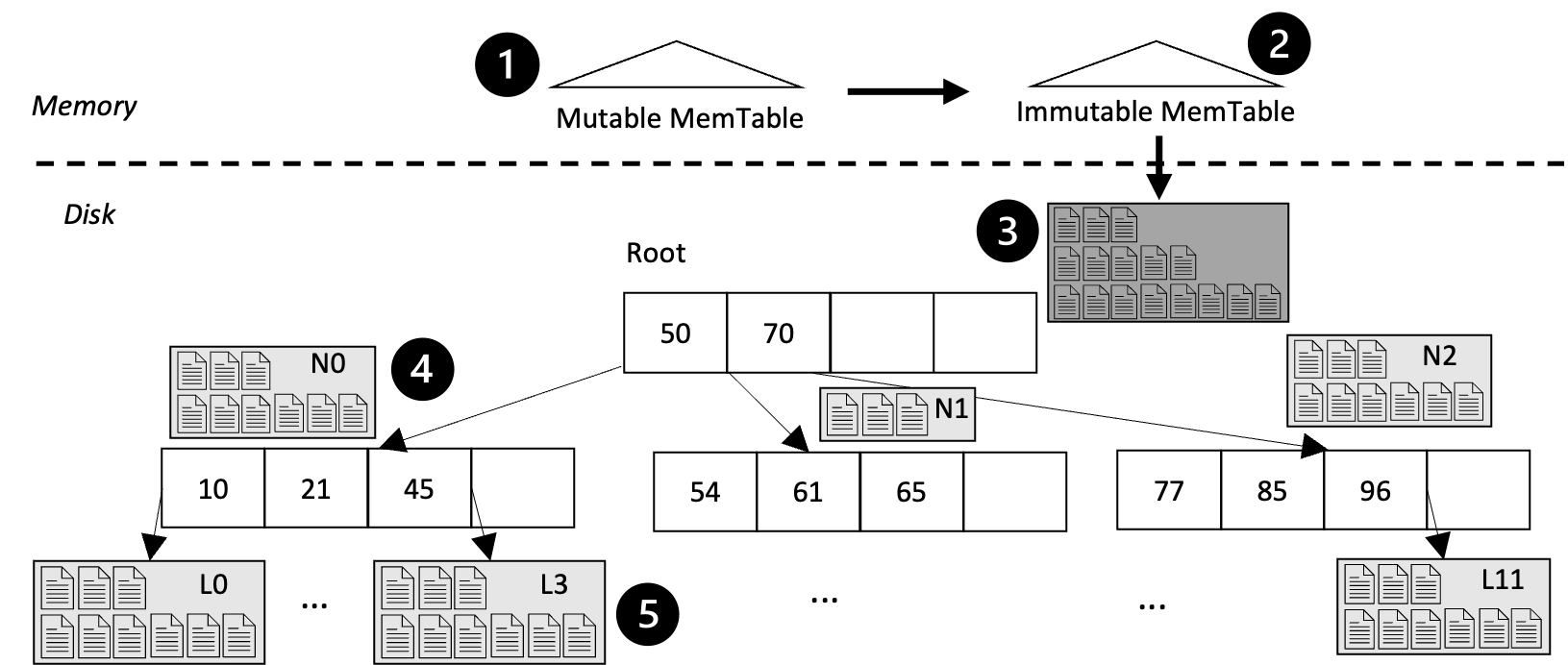}
        \caption{The range search process in \abbrindex{}.}
        \label{fig:scan}
    \end{subfigure}
    \caption{Overview of \abbrindex{}}\label{fig:fig2}
\end{figure*}

New data-intensive applications impose new requirements on data  systems. The term {\em Hybrid Transactional and Analytical Processing} (HTAP, for short) is coined to represent the ever-increasing need to address hybrid requirements, mainly to offer online analytics over online transactional data. Typical HTAP workloads include heavy transactional workloads at times as well as heavy analytical queries at other times. HTAP systems need to support workload changes over time. For example, this is observed as a diurnal pattern in one of the RocksDB use cases at Meta~\cite{cao2020characterizing}, and is  also observed in~\cite{daghistani2021swarm} where the number of tweets fluctuate across the day.

In this demo, we demonstrate a data index termed the Adaptive Hotspot-Aware tree index, (the \abbrindex{}, for short) that is suitable for HTAP workloads. The \abbrindex{} can adapt and morph from being write-optimized at times, e.g., behave like an LSM-tree, to being read-optimized at other times, e.g., behave like a \bplustree{}. We stress-test the \abbrindex{} over one of the most observed changing workloads;  one that oscillates between being write-heavy at times to being read-heavy at other times. This pattern is  practical in some applications, e.g., in a social media app, where users post comments actively during the day, and browse content at night. This corresponds to the oscillating write-heavy posting comments, and read-heavy browsing workloads, which is repeated daily. Another observation over real datasets is that operations are focused on hotspots. Thus, the\abbrindex{} only adapts the structure where hotspot data is stored, which is more time efficient.

Among the analytical workloads, we focus on simple range search queries as we can control the amount of selected data  by the query.
Traditional static indexes, e.g., the \bplustree{}~\cite{bayer1970organization,comer1979ubiquitous} and Log-Structured Merge Tree (LSM-Tree)~\cite{o1996log} that are optimized for only one operation are either optimized for range search query or optimized for write operations, respectively. They cannot achieve the same throughput performance in the case of an oscillating write-heavy and read-heavy workloads.

The ideal index for the oscillating workload is to make the index behave as an LSM-Tree in the write-heavy phase, and behave as a \bplustree{} in the read-heavy workload as  in Figure~\ref{fig:flow}. In this demonstration, we present \abbrindex{} which is an adaptive index being hotspot-aware. We showcase baseline indexes and the \abbrindex{} under oscillating workloads. The \abbrindex{} can adapt itself and catch with the optimal static index within that workload. We also display the intermediate structure of the \abbrindex{} during adapting.  However, as this demo illustrates,  under certain workloads, the overall throughput of the \abbrindex{} is competitive compared to the read-optimized index (i.e., the \bplustree{}) and the write-optimized index (i.e., the LSM-Tree). A trade-off between current performance and future performance is demonstrated for the \abbrindex{}. Meanwhile, being an adaptive index, the \abbrindex{} takes time to adapt. The throughput during the adaptation phase can be low, and this is the tradeoff, but this does not compare to having significant down time to migrate data to suit the new workload or to maintain dual structures, one for each workload type.



\section{Background}
The structure of the \abbrindex{} is inspired by 
the buffer tree~\cite{arge2003buffer}, where 
the \abbrindex{}
can be viewed as a tree part (a \bplustree{}) plus a buffer part (an LSM-Tree), and lies perfectly in the middle 
between a \bplustree{} and an LSM-Tree. However, a buffer tree may not solve the problem of oscillating workloads as it is still a static structure.

The core of adaptation is how to place the buffered data. During a read-heavy workload, buffered data is 
migrated gradually
to the leaf nodes so that range searches 
access 
the index as if it is a \bplustree{}. During the write-heavy workload, data is buffered in batches as if it is an LSM-tree to avoid I/Os caused by individual insertions.

Making the buffer tree adaptive introduces several challenges including: 
\begin{enumerate}
    \item How to query the buffer tree? 
    \item How to maximize the write throughput? 
    \item How to minimize the number of file compactions? 
\end{enumerate}
    We propose techniques to address the above challenges given the observation that real datasets are skewed with hotspots. We apply these techniques in the proposed  \fullindex{}, or \abbrindex{}, for short. We compare the three indexes as  in Figure~\ref{fig:demo}. All indexes are pre-constructed before the first operation. The workload starts from range search queries only, and transitions to update-only when 1 million operations have finished. After another 1 million updates, the workload is transitioned to range search queries only. The \abbrindex{} is able to adapt itself as its throughput climbs during the first workload phase. In the 
    update phase, 
    the 
    \abbrindex{} is competitive with LevelDB. In the third  phase (the read-heavy  range search queries phase), the \abbrindex{} can still adapt itself although finish after more operations than the first phase.

\section{Overview of the \abbrindex{} }\label{section:design}
In this section, we summarize the basic components of the \abbrindex{} as well as how insertions and range search queries are performed in the \abbrindex{}. Then, we briefly explain the adaptation process.

\noindent
{\bf The Index Structure.} The \abbrindex{} combines the LSM-Tree, the \bplustree{}, and the buffer tree. Under a write-heavy workload, all write operations are performed as if the \abbrindex{} is an LSM-Tree. Under a read-heavy workload within a hotspot, all range  search  queries are as if the \abbrindex{} is a \bplustree{} (Figure~\ref{fig:flow}). A key challenge is how to maintain a valid index structure. We use a buffer tree~\cite{arge2003buffer}-like structure as the intermediate structure. The buffer of the root node (\rtl{}) is an LSM-Tree that accepts all the incoming writes. When a buffer overflows, data items are dispatched to the buffers of the children. The buffer of the leaf node (\ndl{}) is also an LSM-Tree. The \abbrindex{} keeps the invariant that \textit{data in \rtl{} are fresher than data in \ndl{}; the closer \ndl{} to the root, the fresher the data}. This invariant is enforced at all times.

\noindent
{\bf Insert.}
Updates are first buffered in the root node's  \rtl{} \verb|MemTable|. 
Once this buffer becomes full,
data items are written to disk in a persistent file, termed an \verb|SSTable| (MemTables and SSTables are as used in LevelDB~\cite{leveldb} and RocksDB~\cite{rocksdb}). An \verb|SSTable| is added to level-0 of \rtl{}. In Figure~\ref{fig:insert}, Step~1 shows the transformation from an LSM-Tree to the construction of the initial tree structure. Both the \rtl{} and \ndl{} have a size limit that when reached for a \rtl{} or \ndl{} of a non-leaf node, a \textit{level-emptying process} is triggered. If \ndl{} of a leaf node reaches the size limit, it is split to multiple leaf nodes each with one \ndl{}. New routing keys are added to the parent of this leaf node, and this may be propagated up til the root.

\noindent
\textbf{Range Search.}
Unlike the buffer tree~\cite{arge2003buffer}, range search queries in the \abbrindex{} are not batched because we want to minimize the latency of individual queries. Since the requested date items may reside in all \ndl{} and leaf pages that have a range that overlaps the query, we rely on a \textit{merged iterator} to produce sorted results. In Figure~\ref{fig:scan},  Range Query [45, 49] starts from the \verb|MemTable| and to \rtl{} that has the freshest data on disk. Then, the query follows the link in the tree page, and finds the appropriate data in \ndl{} of N0 as well as Leaf Node L3.

\noindent
\textbf{Adapting.}
 \abbrindex{} starts to adapt when there is a range search query based on a forming hotspot. The range query records the touched nodes, and sends them to the background thread. To adapt, data of \ndl{}s that are inside the hotspot are flushed down to the leaf level, and the leaf nodes are transformed into leaf pages.

\section{Demonstration Scenario}

\begin{figure}[h]
    \centering
    \includegraphics[width=\linewidth]{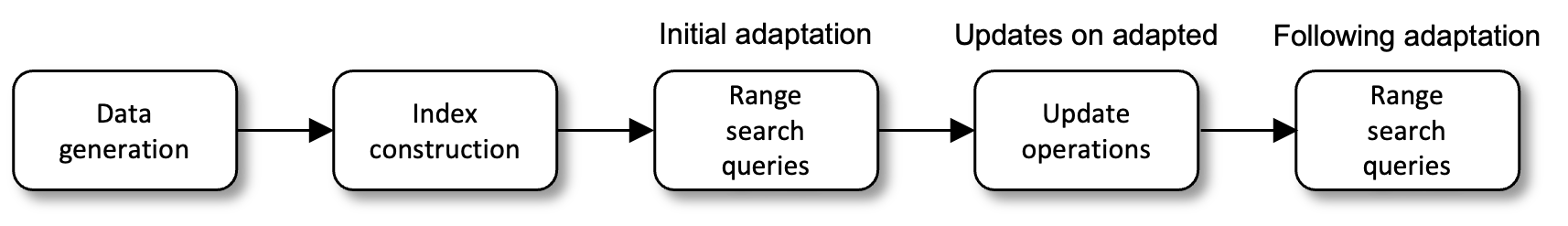}
    \caption{The workflow of the demonstration}
    \label{fig:workflow}
\end{figure}

Figure~\ref{fig:workflow} gives the workflow of our demo that constitutes  data generation, index construction, initial adaptation upon the first range search phase, updates on adapted \abbrindex{} during the update phase, as well as the following range search phase. We will demonstrate, via visualization, the evolution of the \abbrindex{} for alternating write- and read-heavy workloads. We will show how the \abbrindex{} 
adapts
in the hotspot region while the other parts of the structure remain unchanged. We provide several knobs s.t. users can visualize the evolution and performance (online throughput and latency) of the \abbrindex{} under various workload conditions. 

We will contrast the performance of the \abbrindex{} against the \bplustree{} and the LSM-tree under the same conditions. We will provide the users with a simulation knob to identify the location of the hotspot. Users will visualize the simulation of how the index evolves along with its performance. By switching to a read-only workload, we will show how the \abbrindex{} performs during the adaptation process and how other two indexes stay static, as well as how the \abbrindex{} eventually performs after the adaptation process is complete as in Figure~\ref{fig:demo}. Upon switching to a write-only workload again, we will show the \abbrindex{}'s throughput and demonstrate the trade-offs between different adaptation techniques.

\begin{figure*}[h]
    \centering
    \includegraphics[width=0.9\textwidth]{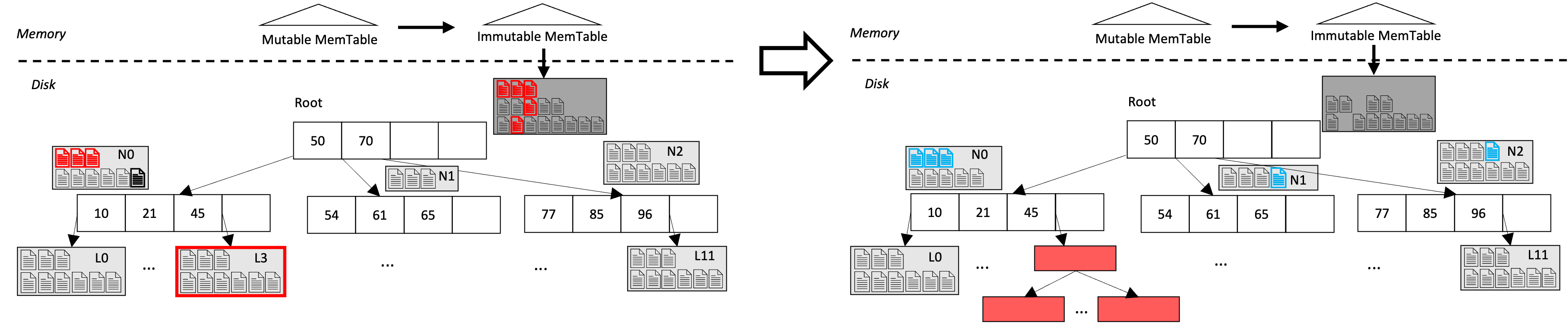}
    \caption{Example of a resulting \abbrindex{} with hotspot region from 46 to 50 after hotspot is fully adapted.}
    \label{fig:w2s}
\end{figure*}

\begin{figure*}[h]
    \begin{subfigure}{0.49\linewidth}
    \centering
        \includegraphics[width=\linewidth]{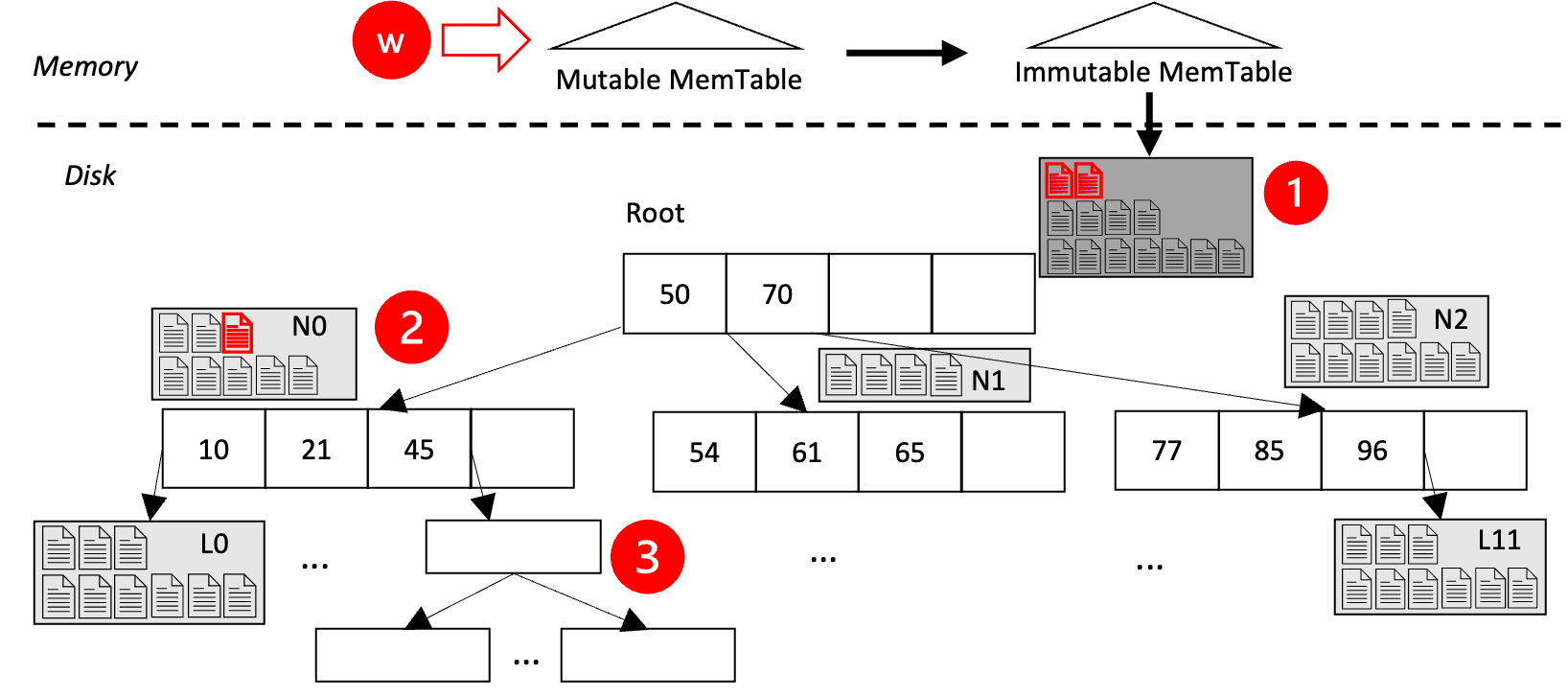}
        \caption{Batched insertion}
        \label{fig:batched-insertion}
    \end{subfigure}
    \hfill
    \begin{subfigure}{0.49\linewidth}
    \centering
        \includegraphics[width=\linewidth]{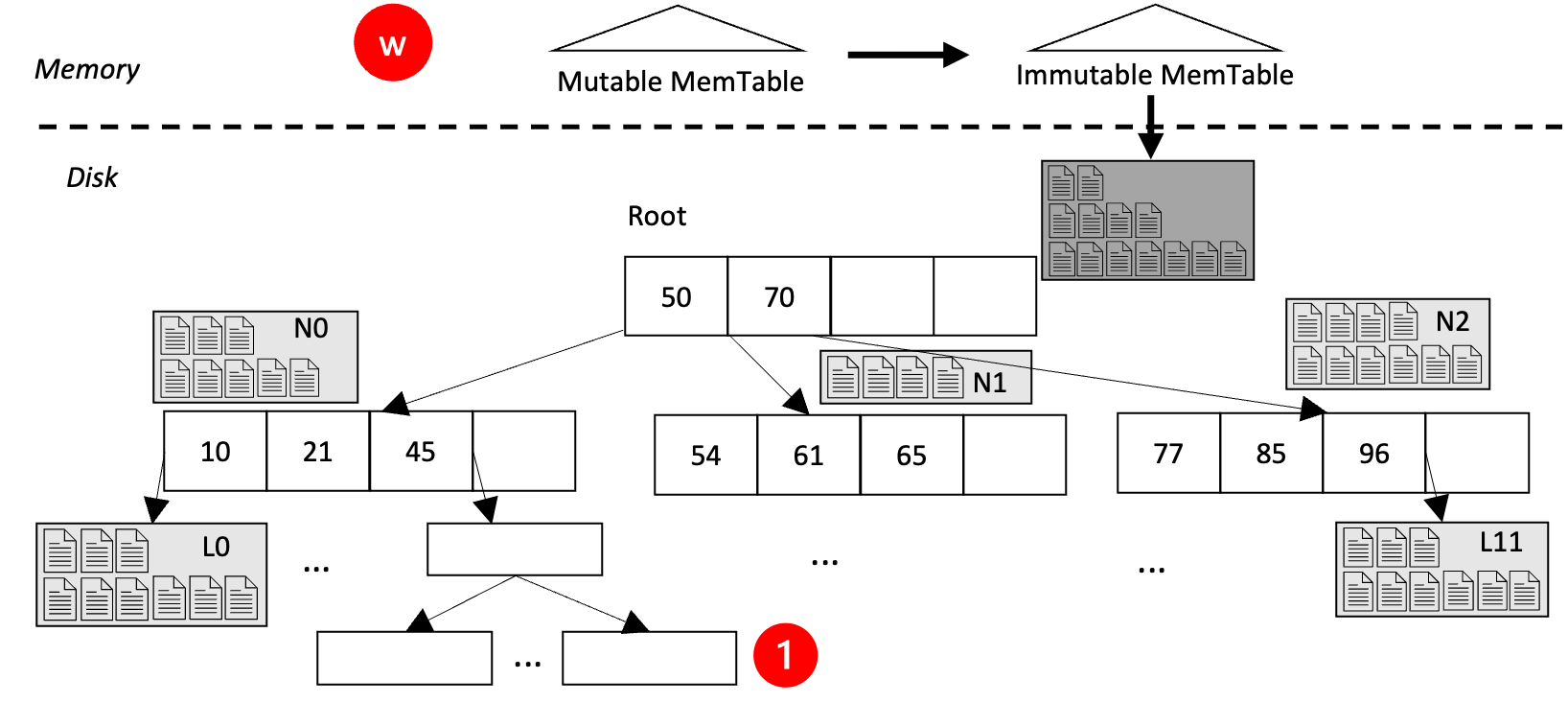}
        \caption{Single insertion}
        \label{fig:single-insertion}
    \end{subfigure}
    \caption{The comparison between batched insertion and single insertion.}\label{fig:fig5}
\end{figure*}

\noindent
\textbf{Knob 1: Data distribution and 
analytics data size.}
The demonstration starts by generating artificial datasets in uniform or Zipfian distributions. Data is loaded into the indexes, including the \abbrindex{}, the \bplustree{} and the LSM-Tree. Users will visualize that how different data distributions may have an effect on the adaptation outcome, and the result after changing the skewness of the Zipfian distribution. In the demonstration, the data size involved in the analytics process is determined by changing the size of the range search. We change the size of the range search as well as make the size dynamically, and show a comparison of the throughputs and latencies among the three index alternatives.

\noindent
\textbf{Knob 2: Enabling/Disabling Adaptation.}
By toggling between enabling and disabling adaptation, users can observe the throughput difference between the two scenarios and the comparison with the baseline indexes, e.g., the \bplustree{} and the LSM-Tree.

\noindent
\textbf{Knob 3: Enabling/Disabling \textit{Seek Compaction}.}
LevelDB's LSM-tree offers  \textit{seek compaction} 
to enhance read latency. Seek compaction
 is triggered after multiple seeks on one file. We present how this technique performs compared to the adapting \abbrindex{}.

\noindent
\textbf{Knob 4: Lazy vs. Eager Adaptation.}
Adaptation can be triggered by each individual range query lazily or by one range query that encloses the entire hotspot region. Typically,  the hotspot location can be predicted beforehand. Figure~\ref{fig:w2s} shows the result after adapting for Hotspot Region [46, 50]. To answer a range query, the \abbrindex{} is traversed top down in all \ndl{}s, starting at \rtl{} and moving to the child node that overlaps  [46, 50], finally reaching the leaf node. Hotspot data is marked in red in the left panel. All the \ndl{}s are recorded, and are sent to the background thread.

The resulting tree has a hybrid structure as in Figure~\ref{fig:w2s} right panel, with the hotspot in a \bplustree{}-like structure, and the other \ndl{}s almost unchanged. Since during hotspot data compaction and flush down, cold spot data that coexists in the original files gets flushed down as well, and is marked in blue in Figure~\ref{fig:w2s} right panel.

In Eager Adaptation, hotspot range [46, 50] is issued as a range query beforehand. In contrast, lazy adaptation relies on the range queries in the workload. Users will be able to visualize the index structure after  full adaptation. Users will also observe the throughput over time of lazy and eager adaptation.


\noindent
\textbf{Knob 5: Balanced vs. Unbalanced Leaf Transformation.}
To adapt a leaf node with a \ndl{} into multiple leaf pages, the straightforward way is to re-compact all the files in \ndl{} and write data items into new pages. However, this incurs re-compaction overhead.
We will demonstrate another way to transform leaf nodes that results in an unbalanced leaf structure as  in Figure~\ref{fig:w2s}.
This relies on the fact that the Level-$i$ ($i > 0$) of the LSM-Tree has sorted files. The transformation takes two steps. First, each file of Level-$i$ ($i > 0$) of the \ndl{} is assigned to a new leaf node. Next, the new node with only one file is split into multiple leaf pages as in Figure~\ref{fig:w2s}.

\noindent
\textbf{Knob 6: Batched vs. Single Inserts.}
After \abbrindex{} has been fully adapted, future inserts may 
insert hotspot data directly into the leaf pages as in Figure~\ref{fig:single-insertion}. Inserts in the cold spot region are still batched into the LSM-tree's memory components before being flushed to the the \ndl{}s.
On the other hand, batched inserts are still batched in memory as in Figure~\ref{fig:batched-insertion}. This raises a trade-off between current  and future performance. Batched inserts are more write-optimized and behave as an LSM-Tree except that Step 3 (See Figure~\ref{fig:batched-insertion}), requires merging files with leaf pages. The 
overhead takes place while adapting during the next read-heavy phase.
While for single insertion, price is paid during write operations and no more adaptation is needed at later range search phase. Users will observe the throughput differences between the two choices.

\noindent
\textbf{Knob 7: Even vs. Sound-remedy Assignment.}
When the leaf node with \ndl{}s are written into leaf pages, the straightforward way is to evenly assign data items across multiple pages. However, this results in \textit{waves of misery}~\cite{glombiewski2019waves,xing2021experimental} and later range queries may suffer depending on the number of update operations. Without mitigating this issue, the \abbrindex{} may experience throughput degradation when adapting back to a read-heavy workload.
To handle this issue, we pack the leaf pages in the initial adaptation phase according to the sound remedy policy introduced in~\cite{glombiewski2019waves,xing2021experimental}. Later in the update phase, hot data are inserted one-by-one using single inserts that prevents waves of misery from happening. We will demonstrate the differences in page utilization distribution between these two assignments. We will also show the throughput with and without the Sound Remedy policy.

\section{Related Work}\label{section:related}
Numerous studies have been conducted in improving the write performance of tree-like structures~\cite{arge2003buffer, graefe2004write, bender2015introduction, zeighami2019nested}. Research has been conducted to improve the LSM-Tree's read performance, including using Bloom filters~\cite{bloom1970space} to improve LSM-Tree point reads, using a hierarchically-stacked Bloom filters to improve range search~\cite{luo2020rosetta} and using a sorted view across multiple files to improve range search~\cite{zhong2021remix}. There are numerous research studies for adaptive indexes, e.g., database cracking~\cite{idreos2007database}, adaptive hybrid indexes~\cite{anneser2022adaptive}, VIP-hashing~\cite{kakaraparthy2022vip}, SA-LSM~\cite{zhang2022sa}, LASER~\cite{saxena2023real}, B$^{link}$-hash~\cite{cha2023blink} among others.

\begin{acks}
Walid G. Aref acknowledges the support of the National Science Foundation under Grant Number IIS-1910216.
\end{acks}


\bibliographystyle{ACM-Reference-Format}
\bibliography{sample}

\end{document}